\documentclass[superscriptaddress,twocolumn,showpacs,preprintnumbers,floatfix]{revtex4}

\usepackage{url}
\usepackage{graphics}
\usepackage{amsfonts}
\usepackage{amsmath}
\usepackage{amssymb}
\usepackage{epsf}

\newcommand{\eM}     {$\epsilon$-machine}
\newcommand{\eMs}    {$\epsilon$-machines}

\newcommand{\EMs}    {$\epsilon$-Machines}

\newcommand{\BiInfinity}	{ \stackrel{\leftrightarrow} {S} }

\newcommand{\Prob}		{ {\rm P}}

\newcommand{\hmu}		{ {h_\mu}}
\newcommand{\EE}		{ {\bf E}}

\newcommand{\mc}[1]{\mathcal{#1}}

\begin{document}

\title{The Organization of Intrinsic Computation:\\
Complexity-Entropy Diagrams and\\
the Diversity of Natural Information Processing}

\author{David P. Feldman}
\email{dave@hornacek.coa.edu}
\affiliation{College of the Atlantic, Bar Harbor, MA 04609}
\affiliation{Santa Fe Institute, 1399 Hyde Park Road, Santa Fe, NM 87501}
\affiliation{Complexity Sciences Center and Physics Department,
University of California, Davis, One Shields Ave, Davis CA 95616}

\author{Carl S. McTague}
\email{c.mctague@dpmms.cam.ac.uk}
\affiliation{DPMMS, Centre for Mathematical Sciences,
University of Cambridge, Wilberforce Road, Cambridge, CB3 0WB, England}
\affiliation{Santa Fe Institute, 1399 Hyde Park Road, Santa Fe, NM 87501}

\author{James P. Crutchfield}
\email{chaos@cse.ucdavis.edu}
\affiliation{Complexity Sciences Center and Physics Department,
University of California, Davis, One Shields Ave, Davis CA 95616}
\affiliation{Santa Fe Institute, 1399 Hyde Park Road, Santa Fe, NM 87501}

\date{\today}

\begin{abstract}
Intrinsic computation refers to how dynamical systems store,
structure, and transform historical and spatial information. By
graphing a measure of structural complexity against a measure of
randomness, complexity-entropy diagrams display the range and
different kinds of intrinsic computation across an entire class of
system.  Here, we use complexity-entropy diagrams to analyze intrinsic
computation in a broad array of deterministic nonlinear and linear
stochastic processes, including maps of the interval, cellular
automata and Ising spin systems in one and two dimensions, Markov
chains, and probabilistic minimal finite-state machines. Since
complexity-entropy diagrams are a function only of observed
configurations, they can be used to compare systems without reference
to system coordinates or parameters. It has been known for some time
that in special cases complexity-entropy diagrams reveal that high
degrees of information processing are associated with phase
transitions in the underlying process space, the so-called ``edge of
chaos''. Generally, though, complexity-entropy diagrams differ
substantially in character, demonstrating a genuine diversity of
distinct kinds of intrinsic computation. 
\end{abstract}


\preprint{Santa Fe Institute Working Paper 08-06-XXX}
\preprint{arxiv.org:0806.XXXX [nlin.CD]}


\maketitle

\bibliographystyle{unsrt}



{\bf
Discovering organization in the natural world is one of science's central
goals. Recent innovations in nonlinear mathematics and physics, in concert
with analyses of how dynamical systems store and process information,
has produced a growing body of results on quantitative ways to measure
natural organization. These efforts had their origin in earlier investigations
of the origins of randomness. Eventually, however, it was realized that
measures of randomness do not capture the property of organization. This
led to the recent efforts to develop measures that are, on the one hand,
as generally applicable as the randomness measures but which, on the other,
capture a system's complexity---its organization, structure, memory, regularity,
symmetry, and pattern. Here---analyzing processes from dynamical systems,
statistical mechanics, stochastic processes, and automata theory---we
show that measures of structural complexity are a necessary and useful
complement to describing natural systems only in terms of their randomness.
The result is a broad appreciation of the kinds of information processing
embedded in nonlinear systems. This, in turn, suggests new physical substrates
to harness for future developments of novel forms of computation.
}

\section{Introduction}

The past several decades have produced a growing body of work on ways
to measure the organization of natural systems. (For early work, see,
e.g., Refs.~\cite{Crut83a,Shaw84,Wolf84,Benn86,Hube86,Gras86,Szep86,Erik87,Kopp87,Land88a,Lloy88,Lind88b,Szep89a,Crut89,Benn90,Crut90,Badi91a,Li91,Crut92c,Bate93a};
for more recent reviews, see
Refs.~\cite{Wack94,Ebel97b,Badi97,Feld98a,Feld98b,Bial01a,Crut03a,Shal01a}.)
The original interest derived from explorations, during the 60's to
the mid-80's, of behavior generated by nonlinear dynamical systems.
The thread that focused especially on pattern and structural complexity
originated, in effect, in attempts to reconstruct geometry \cite{Pack80},
topology \cite{Muld93a}, equations of motion \cite{Crut87a}, periodic orbits
\cite{Auer87a}, and stochastic processes \cite{Fras90b} from observations
of nonlinear processes.  More recently, developing
and using measures of complexity has been a concern of researchers
studying neural computation \cite{Tono94a,Wenn05a}, the clinical
analysis of patterns from a variety of medical signals and imaging
technologies \cite{Sapa98a,Marw02a,Youn05a}, and machine learning and
synchronization \cite{Bial00a,Neme00a,Crut01b,Debo04a,Feld04a},
to mention only a few contemporary applications.

These efforts, however, have their origin in an earlier period in
which the central concern was not the emergence of organization,
but rather the origins of randomness. Specifically, measures were
developed and refined that quantify the degree of randomness and
unpredictability generated by dynamical systems. These
quantities---metric entropy, Lyapunov characteristic exponents,
fractal dimensions, and so on---now provide an often-used and well
understood set of tools for detecting and quantifying deterministic
chaos of various kinds. In the arena of stochastic processes,
Shannon's entropy rate predates even these and has been productively
used for half a century as a measure of an information source's degree
of randomness or unpredictability \cite{Cove91}. 

Over this long early history, researchers came to appreciate that
dynamical systems were capable of an astonishing array of behaviors
that could not be meaningfully summarized by the entropy rate or
fractal dimension.  The reason for this is that, by their definition,
these measures of randomness do not capture the property of
organization. This realization led to the considerable contemporary
efforts just cited to develop measures that are as generally
applicable as the randomness measures but that capture a system's
complexity---its organization, structure, memory, regularity,
symmetry, pattern, and so on.  

Complexity measures which do this are often referred to as {\em
statistical} or {\em structural complexities} to indicate that they
capture a property distinct from randomness. In contrast, {\em
deterministic complexities}---such as the Shannon entropy rate,
Lyapunov characteristic exponents, and the Kolmogorov-Chaitin
complexity---are maximized for random systems. In essence, they are
simply alternatives to measuring the same property---degrees of
randomness. Here, we shall emphasize complexity of the structural
and statistical sort which measures a property complementary to
randomness.  We will demonstrate, across a broad range of model
systems, that measures of structural complexity are a necessary and
useful addition to describing a process in terms of its randomness.

\subsection{Structural Complexity}

How might one go about developing a structural complexity measure? A typical
starting point is to argue that that the structural complexity of a system
must reach a maximum between the system's perfectly ordered and perfectly
disordered extremes
\cite{Crut82b,Hube86,Gras86,Benn90,Crut89,Crut92c,Kopp87,Gell96}.
The basic idea behind these claims is that a system which is either
perfectly predictable (e.g., a periodic sequence) or perfectly
unpredictable (e.g., a fair coin toss) is deemed to have zero
structural complexity. Thus, the argument goes, a system with either
zero entropy or maximal entropy (usually normalized to one), has zero
complexity; these systems are simple and not highly structured. This
line of reasoning further posits that in between these extremes lies
complexity.  Those objects that we intuitively consider to be complex
must involve a continuous element of newness or novelty (i.e.,
entropy), but not to such an extent that the novelty becomes
completely unpredictable and degenerates into mere noise.  

In summary, then, it is common practice to require that a structural
complexity measure vanish in the perfectly ordered and perfectly
disordered limits. Between these limits, the complexity is usually
assumed  to achieve a maximum.  These requirements are often
taken as axioms from which one constructs a complexity measure
that is a single-valued function of randomness as measured by, say, entropy.
In both technical and popular scientific literatures, it is not uncommon
to find a ``complexity'' plotted against entropy in merely schematic form
as a sketch of a generic complexity function that vanishes for extreme
values of entropy and achieves a maximum in a middle region
\cite{Hube86,Atla91a,Gell94a,Flak99a}.  Several authors, in fact, have taken
these as the \emph{only} constraints defining complexity
\cite{Shin99,Lope95,Plas96,Calb01a,Lope01a}.

Here we take a different approach: {\em We do not prescribe how
complexity depends on entropy.} One reason for this is that a useful
complexity measure needs to do more than satisfy the boundary
conditions of vanishing in the high- and low-entropy limits
\cite{Feld98a,Crut00a,Bind00}. In particular, a useful complexity
measure should have an unambiguous interpretation that accounts in
some direct way for how correlations are {\em organized} in a system.
To that end we consider a well defined and frequently used
complexity measures---the \emph{excess entropy}---and empirically
examine its relationship to entropy for a variety of systems.

\subsection{Complexity-Entropy Diagrams}

The diagnostic tool that will be the focal point for our studies is
the {\em complexity-entropy diagram}. Introduced in
Ref.~\cite{Crut89}, a complexity-entropy diagram plots structural
complexity (vertical axis) versus randomness (horizontal axis) for
systems in a given model class. Complexity-entropy diagrams
allow for a direct view of the complexity-entropy relationship within
and across different systems. For example, one can easily read whether
or not complexity is a single-valued function of entropy.

The complexity and entropy measures that we use capture a system's
{\em intrinsic computation} \cite{Crut92c}: how a system stores,
organizes, and transforms information.  A crucial point is that these
measures of intrinsic computation are properties of the system's
configurations. They do not require knowledge of the equations of
motion or Hamiltonian or of system parameters (e.g., temperature,
dissipation, or spin-coupling strength) that generated the configurations.
Hence, in addition to the many cases in which they can be calculated
analytically, they can be inductively calculated from observations of
symbolic sequences or configurations. 

Thus, a complexity-entropy diagram measures intrinsic computation
in a parameter-free way.  This allows for the direct comparison of
intrinsic computation across very different classes since a
complexity-entropy diagram expresses this in terms of common
``information-processing'' coordinates.  As such, a complexity-entropy
diagram demonstrates how much a given resource (e.g., stored
information) is required to produce a given amount of randomness
(entropy), or how much novelty (entropy) is needed to produce a
certain amount of statistical complexity. 

Recently, a form of complexity-entropy diagram has been used in the
study of anatomical MRI brain images \cite{Youn05a,Youn08a}.  This
work showed that complexity-entropy diagrams give a reliable way
to distinguish between ``normal'' brains and those experiencing
cortical thinning, a condition associated with Alzheimer's disease.
Complexity-entropy diagrams have also recently been used as part of a
proposed test to distinguish chaos from noise \cite{Ross07a}.  And
Ref.~\cite{Mart06a} calculates complexity-entropy diagrams for a
handful of different complexity measures using the sequences generated
by the symbolic dynamics of various chaotic maps. 

Historically, one of the motivations behind complexity-entropy diagrams was
to explore the common claim that complexity achieves a {\em sharp} maximum
at a well defined boundary between the order-disorder extremes. This led,
for example, to the widely popularized notion of the ``edge of chaos''
\cite{Pack88a,Kauf93a,Lang90a,Wald92a,Ray94a,Melb00a,Bert03a,Bert04a}---namely,
that objects achieve maximum complexity at a \emph{boundary} between
order and disorder. 
Although these particular claims have been criticized \cite{Mitc93}, during
the same period it was shown that at the \emph{onset of chaos} complexity
does reach a maximum. Specifically, Ref.~\cite{Crut89} showed that the
\emph{statistical complexity} diverges at the accumulation point of the
period-doubling route to chaos. This led to an analytical theory that
describes exactly the interdependence of complexity and entropy for this
universal route to chaos \cite{Crut90}. Similarly, another complexity measure,
the \emph{excess entropy} \cite{Crut83a,Shaw84,Gras86,Lind88b,Li91,Crut03a,Rate96a,Freu96a,Schu02a}
has also been shown to diverge at the period-doubling critical point.

This latter work gave some hope that there would be a universal relationship
between complexity and entropy---that some appropriately defined measure of
complexity plotted against an appropriate entropy would have the same
functional form for a wide variety of systems. In part, the motivation for
this was the remarkable success of scaling and data collapse for critical
phenomena.  Data collapse is a phenomena in which certain variables for very
different systems collapse onto a single curve when appropriately rescaled
near the critical point of a continuous phase transition. For example, the
magnetization and susceptibility exhibit data collapse near the
ferromagnet-paramagnet transition.  See, for example,
Refs.~\cite{Stan99a,Yeom92} for further discussion.
Data collapse reveals that different
systems---e.g., different materials with different critical
temperatures---possess a deep similarity despite differences in their details.

The hope, then, was to find a similar universal curve for complexity as
a function of entropy. One now sees that this is not and, fortunately, cannot
be the case. Notwithstanding special parametrized examples, such as
period-doubling and other routes to chaos, a wide range of complexity-entropy
relationships exists \cite{Crut92c,Li91,Crut97a,Feld98a}. This is a point
that we will repeatedly reinforce in the following.

\subsection{Surveying Complexity-Entropy Diagrams}

We will present a survey of the relationships between structure and
randomness for a number of familiar, well studied systems including
deterministic nonlinear and linear stochastic processes and well known
models of computation. The systems we study include maps of the
interval, cellular automata and Ising models in one and two
dimensions, Markov chains, and minimal finite-state machines. To our
knowledge, this is the first such cross-model survey of
complexity-entropy diagrams. 

The main conclusion that emerges from our results is that there is a large
range of possible complexity-entropy behaviors. Specifically, there is
not a universal complexity-entropy curve, there is not a general
complexity-entropy transition, nor is it case that complexity-entropy
diagrams for different systems are even qualitatively similar. These results
give a concrete picture of the very different types of relationship between
a system's rate of information production and the structural organization
which produces that randomness. This diversity opens up a number
of interesting mathematical questions, and it appears to suggest a new
kind of richness in nature's organization of intrinsic computation.

Our exploration of intrinsic computation is structured as follows: In
Section \ref{Info.Theory.Review} we briefly review several
information-theoretic quantities, most notably the entropy rate and
the excess entropy. In Section \ref{Comp.Ent.Section} we present results
for the complexity-entropy diagrams for a wide range of model systems.
In Section \ref{Discussion} we discuss our results, make a number
of general comments and observations, and conclude by summarizing.

\section{Entropy and Complexity Measures}
\label{Info.Theory.Review}

\subsection{Information-Theoretic Quantities}

The complexity-entropy diagrams we will examine make use of two
information-theoretic quantities: the excess entropy and the entropy
rate.  In this section we fix notation and give a brief but
self-contained review of them.

We begin by describing the stochastic process generated by a system.
Specifically, we are interested here in describing the character of
bi-infinite, one-dimensional sequences:
$\BiInfinity = \ldots,  S_{-2}, S_{-1}, S_0, S_1, \ldots$, where
the $S_i$'s are random variables that assume values $s_i$ in a
finite alphabet $\mathcal{A}$. Throughout, we follow the standard
convention that a lower-case letter refers to a particular value of
the random variable denoted by the corresponding upper-case letter.
In the following, the index $i$ on the $S_i$'s will refer to either
space or time.

A \emph{process} is, quite simply, the distribution over all possible
sequences generated by a system: $\Prob(\BiInfinity)$. Let $\Prob(s_i^L)$
denote the probability that a block $S_i^L = S_i S_{i+1} \ldots S_{i+L-1}$
of $L$ consecutive symbols takes on the particular values
$s_i, s_{i+1}, \ldots , s_{i+L-1} \in \mathcal{A}$. We will assume that
the distribution over blocks is stationary: $\Prob(S_i^L) =
\Prob(S_{i+M}^L)$ for all $i$, $M$, and $L$. And so we will drop the
index on the block probabilities.  When there is no confusion, then,
we denote by $s^L$ a particular sequence of $L$ symbols, and use
$\Prob(s^L)$ to denote the probability that the particular $L$-block
occurs. 

The \emph{support} of a process is the set of allowed sequences---i.e.,
those with positive probability. In the parlance of computation theory,
a process' support is a formal language: the set of all finite length
words that occur at least once in an infinite sequence.

A special class of processes that we will consider in subsequent
sections are {\em Order-$R$ Markov Chains}.  These processes are those
for which the joint distribution can be conditionally factored into
words $S^R$ of length $R$---that is, 
\begin{equation}
\Prob(\BiInfinity) = \ldots \Prob(S_i^R|S_{i-R}^R)
  \Prob(S_{i+R}^R|S_i^R) \Prob(S_{i+2R}^R|S_{i+R}^R) \ldots \;.
\label{R.Markovian}
\end{equation}
In other words, knowledge of the current length-$R$ word is all that
is needed to determine the distribution of future symbols.  As a
result, the states of the Markov chain are associated with the
$\mathcal{A}^R$ possible values that can be assumed by a length-$R$
word. 

We now briefly review several central quantities of information theory
that we will use to develop measures of unpredictability and entropy.
For details see any textbook on information theory; e.g.,
Ref.~\cite{Cove91}.  Let $X$ be a random variable that assumes the
values $x \in {\cal X}$, where ${\cal X}$ is a finite set. The probability
that $X$ assumes the value $x$ is given by $\Prob(x)$. Also, let $Y$
be a random variable that assumes values $y\in {\cal Y}$.

The \emph{Shannon entropy} of the variable $X$ is given by:
\begin{equation}
H[X] \, \equiv \,  - \sum_{x \in {\cal X}} \Prob(x) \log_2 {\rm
P}(x) \; .
\end{equation}
The units are given in \emph{bits}. This quantity measures the uncertainty
associated with the random variable $X$. Equivalently, $H[X]$ is also
the average amount of memory needed to store outcomes of variable $X$.

The \emph{joint entropy} of two random variables, X and Y, is defined as:
\begin{equation}
H[X,Y] \, \equiv \, -\sum_{x \in {\cal X}, y \in {\cal Y} } \Prob(x,y)
\log_2 \Prob(x,y) \;.
\end{equation}
It is a measure of the uncertainty associated with the joint distribution
$\Prob(X,Y)$. The \emph{conditional entropy} is defined as:
\begin{equation}
H[X|Y]  \, \equiv \,- \sum_{x \in {\cal X}, y \in {\cal Y} }
\Prob(x,y) \log_2  \Prob(x|y) \;,
\end{equation}
and gives the average uncertainty of the conditional probability
$\Prob(X|Y)$. That is, $H[X|Y]$ tells us how uncertain, on
average, we are about $X$, given that the outcome of $Y$ is known.

Finally, the \emph{mutual information} is defined as:
\begin{equation}
I [X;Y]  \, \equiv \, H[X] - H[X|Y] \;.
\label{MI.def}
\end{equation}
It measures the average reduction of uncertainty
of one variable due to knowledge of another.  If knowing $Y$ on
average reduces uncertainty about $X$, then it makes sense to say that
$Y$ carries information about $X$.  Note that $I[X;Y] = I[Y;X]$.

\subsection{Entropy Growth and Entropy Rate}

With these definitions set, we are ready to develop an information-theoretic
measure of a process's randomness. Our starting point is to consider blocks
of consecutive variables. The \emph{block entropy} is the total Shannon
entropy of length-$L$ sequences:
\begin{equation}
H(L) \, \equiv \, - \sum_{ s^L \in {\cal A}^L} \Prob(s^L)
  \log_2 \Prob(s^L) \;,
\label{H.def}
\end{equation}
where $L > 0$. The sums run over all possible blocks of
length $L$. We define $H(0) \equiv 0$. The block entropy grows monotonically
with block length: $H(L) \geq H(L-1)$.

For stationary processes the total Shannon entropy typically grows linearly
with $L$.  That is, for sufficiently large $L$, $H(L) \sim L$. This leads
one to define the \emph{entropy rate} $\hmu$ as:
\begin{equation}
   \hmu \, \equiv \,\lim_{L \rightarrow \infty} \frac{H(L)}{L }\;.
\label{hmu.def}
\end{equation}
The units of $\hmu$ are \emph{bits per symbol}.
This limit exists for all stationary sequences \cite[Chapter
  4.2]{Cove91}. The entropy rate is also know as the \emph{metric
  entropy} in dynamical systems theory and is equivalent to the
\emph{thermodynamic entropy density} familiar from equilibrium
statistical mechanics. 

The entropy rate can be given an additional interpretation as
follows.  First, we define an $L$-dependent entropy rate estimate:
\begin{eqnarray}
\hmu(L)  &\, = \,& H(L) - H(L\!-\!1) \\
 & \, = \, & H[S_L|S_{L-1}, S_{L-2} , \ldots , S_1] \;, \;\; L >0 \;.
\label{hmu.L.def}
\end{eqnarray}
We set $\hmu(0) = \log_2 |{\cal A}|$. In words, then, $\hmu(L)$
is the average
uncertainty of the next variable $S_L$, given that the previous
$L\!-\!1$ symbols have been seen.  Geometrically, $h_{\mu}(L)$ is the
two-point slope of the total entropy growth curve $H(L)$.  Since
conditioning on more variables can never increase the entropy, it
follows that $\hmu(L) \leq \hmu(L-1)$.  In the $L \rightarrow \infty$
limit, $\hmu(L)$ is equal to the entropy rate defined above in
Eq.~(\ref{hmu.def}):
\begin{equation}
  \hmu \, = \, \lim_{L \rightarrow \infty} \hmu(L) \;.
\label{hmu.conditional}
\end{equation}
Again, this limit exists for all stationary processes \cite{Cove91}.
Equation (\ref{hmu.conditional}) tells us that $\hmu$ may be viewed
as the irreducible randomness in a process---the randomness
that persists even after statistics over longer and longer blocks of
variables are taken into account.

\subsection{Excess Entropy}

The entropy rate gives a reliable and well understood measure of
the randomness or disorder intrinsic to a process. However, as the
introduction noted, this tells us little about the underlying system's
organization, structure, or correlations.  Looking at the manner in
which $\hmu(L)$ converges to its asymptotic value $\hmu$, however,
provides one measure of these properties.

When observations only over length-$L$ blocks are taken into account,
a process appears to have an entropy rate of $\hmu(L)$.  This
quantity is larger than the true, asymptotic value of the entropy
rate $\hmu$.  As a result, the process appears more random by
$\hmu(L) - \hmu$ bits.  Summing these entropy over-estimates over
$L$, one obtains the {\em excess entropy}
\cite{Crut83a,Shaw84,Gras86,Lind88b}:
\begin{equation}
\EE \, \equiv \,  \sum_{L=1}^{\infty} [\hmu(L) - \hmu] \;.
\label{E.def}
\end{equation}
The units of $\EE$ are \emph{bits}. The excess entropy tells us
how much information must be gained before it is possible to infer
the actual per-symbol randomness $\hmu$.  It is large if the system
possesses many regularities or correlations that manifest themselves
only at large scales. As such, the excess entropy can serve as a
measure of global structure or correlation present in the system. 

This interpretation is strengthened by noting that the excess
entropy can also be expressed as the mutual information between two
adjacent semi-infinite blocks of variables \cite{Li91,Crut03a}:
\begin{equation}
  {\bf E } \, = \, \lim_{L \rightarrow \infty} I[ S_{-L}, S_{-L+1}
  ,S_{-1}; S_0, S_1, \ldots S_{L-1}] \;.
\label{E.mutual.info}
\end{equation}
Thus, the excess entropy measures one type of the memory of the
system; it tells us how much knowledge of one half of the system
reduces our uncertainty about the other half. If the sequence of
random variables is a time series, then $\EE$ is the amount of
information the past shares with the future.

The excess entropy may also be given a geometric interpretation.
The existence of the entropy rate suggests that $H(L)$ grows
linearly with $L$ for large $L$ and that the growth rate, or
slope, is given by $\hmu$.  It is then possible to show that the
excess entropy is the ``$y$-intercept'' of the asymptotic form for
$H(L)$ \cite{Shaw84,Gras86,Li91,Arno96,Bial00a,Neme00a}:
\begin{equation}
  H(L) \, \sim \, \EE + \hmu L ~,
  \; {\rm as} \; L \rightarrow \infty \;.
\label{H.Scaling.Form}
\end{equation}
Or, rearranging, we have
\begin{equation}
  \EE \, = \, \lim_{L\rightarrow \infty} \left[ H(L) - \hmu L
  \right] \;.
\end{equation}

This form of the excess entropy highlights another interpretation: $\EE$ is
the \emph{cost of amnesia}. If an observer has extracted enough information
from a system (at large $L$) to predict it optimally ($\sim \hmu$), but
suddenly loses all of that information, the process will then appear more
random by an amount $H(L) - \hmu L$.

To close, note that the excess entropy, originally coined in \cite{Crut83a},
goes by a number of different names,
including ``stored information'' \cite{Shaw84}; ``effective
measure complexity'' \cite{Gras86,Lind88b,Lind89a,Erik87,Ebel02a};
``complexity'' \cite{Li91,Arno96}; ``predictive information''
\cite{Bial00a,Neme00a}; and ``reduced R\'enyi entropy of order $1$''
\cite{Csor89a,Kauf91a}. For recent reviews on excess entropy, entropy
convergence in general, and applications of this approach see
Refs.~\cite{Ebel97b,Crut03a,Bial00a}.

\subsection{Intrinsic Information Processing Coordinates}
\label{Sec:ComplexityEntropyDiagram}

In the model classes examined below, we shall take the excess
entropy $\EE$ as our measure of complexity and use the entropy rate
$\hmu$ as the randomness measure. The excess entropy ${\bf E}$ and the
entropy rate $\hmu$ are exactly the two quantities that specify the
large-$L$ asymptotic form for the block entropy
Eq.~(\ref{H.Scaling.Form}). 
The set of all $(\hmu, \EE)$ pairs is thus geometrically equivalent to
the set of all straight lines with non-negative slope and intercept.
Clearly, a line's slope and intercept are independent quantities. Thus,
there is no {\em a priori} reason to anticipate any relationship between
$\hmu$ and $\EE$, a point emphasized early on by Li \cite{Li91}.

It is helpful in the following to know that for binary order-$R$
Markov processes there is an upper bound on the excess entropy:
\begin{equation}
\EE \leq R (1-h_\mu) \;.
\label{EE_hmu_bound}
\end{equation}
We sketch a justification of this result here; for the derivation, see
\cite[Proposition 11]{Crut03a}. First, recall that the excess entropy may
be written as the mutual information between two semi-infinite blocks, as
indicated in Eq.~(\ref{E.mutual.info}).  However, given the process is
order-$R$ Markovian, Eq.~(\ref{R.Markovian}), the excess entropy reduces
to the mutual information between two adjacent $R$-blocks. From
Eq.~(\ref{MI.def}), we see that the excess entropy is the entropy of
an $R$-block minus the entropy of an $R$-block conditioned on its neighboring
$R$-block:
\begin{equation}
 \EE \, = \, H(R) - H[S_i^R|S_{i-R}^R] \;.
\label{E.markov}
\end{equation}
(Note that this only holds in the special case of order-$R$ Markov processes.
It is \emph{not} true in general.)
The first term on the right hand side of Eq.~(\ref{E.markov}) is
maximized when the distribution over the $R$-block is uniform, in
which case $H(R) = R$.  The second term on the right hand side is
minimized by assuming that the conditional entropy of the two blocks
is given simply by $R h_\mu$---i.e., $R$ times the per-symbol entropy
rate $h_\mu$.  In other words, we obtain a lower bound by assuming that
the process is independent, identically distributed over $R$-blocks.
Combining the two bounds gives Eq.~(\ref{EE_hmu_bound}).  

It is also helpful in the following to know that for periodic processes
$\hmu = 0$ (perfectly predictable) and $\EE = \log_2 p$, where $p$ is
the period \cite{Crut03a}. In this case, $\EE$ is the amount of information
required to distinguish the $p$ phases of the cycle.

\subsection{Calculating Complexities and Entropies}
\label{numerical.methods.section}

As is now clear, all quantities of interest depend on knowing sequence
probabilities $\Prob (s^L)$. These can be obtained by direct analytical
approximation given a model or by numerical estimation via simulation.
Sometimes, in special cases, the complexity and entropy can be calculated
in closed form.

For some, but not all, of the process classes studied in the following,
we estimate the various information-theoretic quantities by simulation.
We generate a long sequence, keeping track of the frequency of occurrence
of words up to some finite length $L$. The word counts are stored in
a dynamically generated parse tree, allowing us to go out to $L = 120$
in some cases. We first make a rough estimate of the topological entropy
using a small $L$ value. This entropy determines the sparseness of the
parse tree, which in turn determines how large a tree can be stored in
a given amount of memory. From the word and subword frequencies
$\Prob(s^L)$, one directly calculates $H(L)$ and, thus, $\hmu$ and
$\EE$.  Estimation errors in these quantities are a function of
statistical errors in $\Prob(s^L)$.

Here, we are mainly interested in gaining a general sense of the behavior
of the entropy rate $\hmu$ and the excess entropy $\EE$. And so, for the
purposes of our survey, this direct method is sufficient.  The vast
majority of our estimates are accurate to at least $1\%$. If extremely
accurate estimates are needed, there exist a variety of techniques for
correcting for estimator bias
\cite{Gras88a,Gras89a,Herz94a,Schu96,deWi99a,Neme02a}.  When one is
working with finite data, there is also the question of what errors
occur, since the $L \rightarrow \infty$ limit cannot be taken. For
more on this issue, see Ref.~\cite{Crut03a}. 

Regardless of these potential subtleties, the entropy rate and
excess entropy can be reliably estimated via simulation, given access
to a reasonably large amount of data. Moreover, this estimation is
purely inductive---one does not need to use knowledge of the
underlying equations of motion or the hidden states that produced the
sequence.  Nevertheless, for several of the model classes we
consider---one-dimensional Ising models, Markov chains, and
topological Markov chains---we calculate the quantities using
closed-form expressions, leading to essentially no error.


\section{Complexity-Entropy Diagrams}
\label{Comp.Ent.Section}

In the following sections we present a survey of intrinsic computation
across a wide range of process classes.
We think of a \emph{class} of system as given by equations of motion,
or other specification for a stochastic process,
that are parametrized in some way---a pair of control parameters
in a one-dimensional map or the energy of a Hamiltonian, say. The
space of parameters, then, is the concrete representation of the
space of possible systems, and a class of system is a subset of the
set of all possible processes. A point in the parameter space is then
a particular \emph{system}, whose intrinsic computation we will
summarize by a pair of numbers---one a measure of randomness, the
other a measure of structure. In several cases, these measures are estimated
from sequences generated by the temporal or spatial process.

\subsection{One-Dimensional Discrete Iterated Maps}

Here we look at the symbolic dynamics generated by two iterated maps
of the interval---the well studied \emph{logistic} and \emph{tent
maps}---of the form:
\begin{equation}
x_{n+1} = f_\mu (x_n) ~,
\end{equation}
where $\mu$ is a parameter that controls the nonlinear function $f$,
$x_n \in [0,1]$, and one starts with $x_0$, the \emph{initial condition}.
The logistic and tent maps are canonical examples of systems
exhibiting deterministic chaos.
The nonlinear iterated function $f$  consists of two monotone
pieces. And so, one can analyze the maps' behavior on the interval via
a \emph{generating partition} that reduces a sequence of continuous states
$x_0, x_1, x_2, \ldots$ to a binary sequence $s_0, s_1, s_2,
\ldots$ \cite{Bai89a}. The binary partition is given by
\begin{equation}
s_i =  \left\{
  \begin{array}{cl}
	0 & x \leq \frac{1}{2} \\
	\\
	1 & x > \frac{1}{2}
  \end{array}
  \right. ~.
\end{equation}
The binary sequence may be viewed as a {\em code} for the set of initial
conditions that produce the sequence. When the maps are chaotic, arbitrarily
long binary
sequences produced using this partition code for arbitrarily small
intervals of initial conditions on the chaotic attractor. Hence, one
can explore many of these maps' properties via binary sequences.

\subsubsection{Logistic Map}

\begin{figure}[tbp]
\epsfxsize=3.2in
\begin{center}
\leavevmode
\epsffile{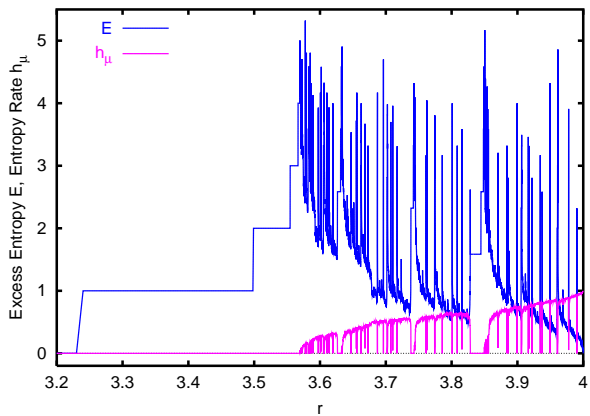}
\end{center}
\vspace{-.6cm}
\caption{Excess entropy $\EE$ and entropy rate $\hmu$ as a function of
the parameter $r$.  The top curve is excess entropy. The $r$ values
were sampled uniformly as $r$ was varied from $3.4$ to $4.0$ in
increments of $0.0001$.  The largest $L$ used was $L=30$ for systems
with low entropy.  For each parameter value with positive entropy, $1
\times 10^7$ words of length $L$ were sampled.  }
\label{logistic.E.hmu.vs.r.plot}
\vspace{-.2cm}
\end{figure}

We begin with the logistic map of the unit interval:
\begin{equation}
    f(x) \, = \, rx(1-x) \;,
\label{logistic}
\end{equation}
where the control parameter $r \in [0,4]$. We iterate this starting
with an arbitrary initial condition $x_0 \in [0,1]$.  In
Fig.~\ref{logistic.E.hmu.vs.r.plot} we show numerical estimates of the
excess entropy $\EE$ and the entropy rate $\hmu$ as a function of
$r$.  Notice that both $\EE$ and $\hmu$ change in a complicated matter
as the parameter $r$ is varied continuously.

As $r$ increases from $3.0$ to approximately $3.5926$,
the logistic map undergoes a series of period-doubling bifurcations.
For $r \in (3.0,3.2361)$ the sequences generated by the logistic
map are periodic with period two, for $r \in
(3.2361,3.4986)$ the sequences are period 4,
and for $r \in (3.4986,3.5546)$ the sequences are
period $8$. For all periodic sequences of period $p$, the entropy
rate $\hmu$ is zero and the excess entropy $\EE$ is $\log_2 p$. So, as
the period doubles, the excess entropy increases by one bit.  This can
be seen in the staircase on the left hand side of
Fig.~\ref{logistic.E.hmu.vs.r.plot}.  At $r \approx
3.5926$, the logistic map becomes chaotic, as evidenced by
a positive entropy rate.  For further discussion of the
phenomenology of the logistic map, see almost any modern textbook on
nonlinear dynamics, e.g., Refs.~\cite{Peit92,Ott93}. 

\begin{figure}[htbp]
\epsfxsize=3.2in
\begin{center}
\leavevmode
\epsffile{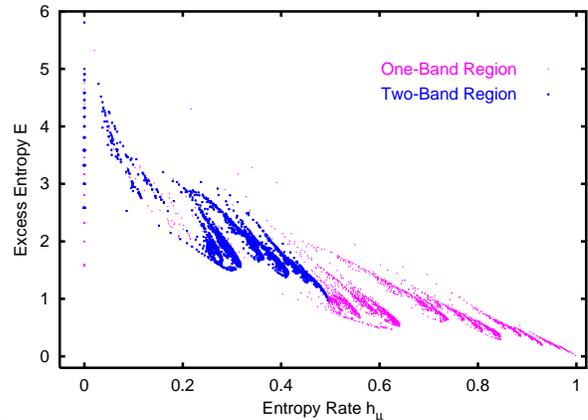}
\end{center}
\vspace{-.8cm}
\caption{Entropy rate and excess entropy $(\hmu,\EE)$-pairs for
  logistic map. Points from regions of the map in which the bifurcation
diagram has one or two bands are colored differently. There are $3214$
parameter values sampled for the one-band region and $3440$ values for
the two-band region.  The $r$ values were sampled uniformly.  The
one-band region is $r \in (3.6786, 4.0)$; the two-band
region is $r \in (3.5926, 3.6786)$. The largest
$L$ used was $L=30$ for systems with low entropy.  For each parameter
value with positive entropy, $1 \times 10^7$ words of length $L$ were
sampled.  }
\label{logistic.banded.plot}
\vspace{-.2cm}
\end{figure}

Looking at Fig.~\ref{logistic.E.hmu.vs.r.plot}, it is difficult to see
how ${\bf E}$ and $h_\mu$ are related.  This relationship can be seen
much more clearly in Fig.~\ref{logistic.banded.plot}, in which we show
the complexity-entropy diagram for the same system.  That is, we plot
$(\hmu,\EE)$ pairs.  This lets us
look at how the excess entropy and the entropy rate are related,
independent of the parameter $r$.


Figure \ref{logistic.banded.plot} shows that there is a definite
relationship between $\EE$ and $\hmu$---one that is not immediately evident
from looking at Fig.~\ref{logistic.E.hmu.vs.r.plot}.  Note, however,
that this relationship is not a simple one.  In particular, complexity
is not a function of entropy: $\EE \neq g(\hmu)$.  For a given value
of $\hmu$, multiple excess entropy values $\EE$ are possible. 

There are several additional empirical observations to extract from
Fig.~\ref{logistic.banded.plot}. First, the shape appears to be
self-similar. This is not at all surprising, given that the logistic
map's bifurcation diagram itself is self-similar.  Second, note the
clumpy, nonuniform clustering of $(\hmu, \EE)$ pairs within the dense
region. Third, note that there is a fairly well defined lower bound.
Fourth, for a given value of the
entropy rate $h_\mu$ there are many possible values for the excess
entropy $\EE$.   However, it appears as if not all $\EE$ values are
possible for a given $h_\mu$.  Lastly, note that there does not appear to be
any phase transition (at finite $h_\mu$) in the complexity-entropy diagram.
Strictly speaking, such a transition does occur, but it does so at
zero entropy rate.  As the period doublings accumulate, the excess entropy
grows without bound.  As a result, the possible excess entropy values
at $h_\mu = 0$ on the complexity-entropy diagram are unbounded.  For
further discussion, see Ref. \cite{Crut90}.

\subsubsection{Tent Map}

\begin{figure}[tbp]
\epsfxsize=3.0in
\begin{center}
\leavevmode
\epsffile{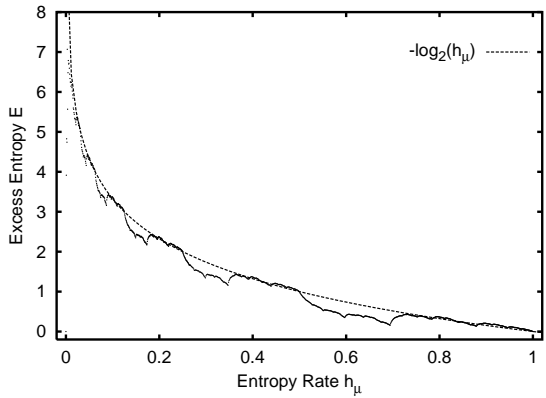}
\end{center}
\vspace{-.6cm}
\caption{Excess entropy $\EE$ versus entropy density $\hmu$ for the
  tent map. The $L$ used to estimate $\Prob(s^L)$, and so $\EE$ and
  $\hmu$, varied depending on the $a$ parameter. The largest $L$ used
  was $L=120$ at low $\hmu$. The plot shows $1,200$ $(\hmu,\EE)$-pairs.
  The parameter was incremented every $\Delta a = 5 \times 10^{-4}$ for
  $a \in [1,1.2]$ and then incremented every $\Delta a = 0.001$ for
  $a \in [1.2,2.0]$. For each parameter value with positive entropy,
  $10^7$ words of length $L$ were sampled.
  }
\label{tent.plot}
\vspace{-.2cm}
\end{figure}

We next consider the {\em tent map}:
\begin{equation}
 f(x) \, = \,  \left\{ \begin{array}{ll} ax &
x < \frac{1}{2} \\ \\ \hspace{1mm}  a(1-x)  & x \geq \frac{1}{2}
        \end{array} \right.   \; ,
\end{equation}
where $a \in [0,2]$ is the control parameter. For $a \in [1,2]$,
the entropy rate $\hmu = \log_2 a$; when $a \in [0,1]$, $\hmu = 0$.
Fig.~\ref{tent.plot} shows $1,200$ $(\hmu,\EE)$-pairs in which
$\EE$ is calculated numerically from empirical estimates of the
binary word distribution $\Prob(s^L)$.

Reference \cite{Crut90} developed a phenomenological theory that explains
the properties of the tent map at the so-called \emph{band-merging points},
where bands of the chaotic attractor merge pairwise as a function of the
control parameter. The behavior at these points is
\emph{noisy periodic}---the order of band visitations is periodic,
but motion within is deterministic chaotic. They occur when
$a = 2^{2^{-n}}$. The symbolic-dynamic process is described by a Markov
chain consisting of a periodic cycle of $2^n$ states in which
all state-to-state transitions are nonbranching except for one where
$s_i = 0$ or $s_i = 1$ with equal probability. Thus, each phase
of the Markov chain has zero entropy per transition, except for the one
that has a branching entropy of $1$ bit. The entropy rate at band-mergings
is thus $\hmu = 2^{-n}$, with $n$ an integer.

The excess entropy for the symbolic-dynamic process at the
$2^n$-to-$2^{n-1}$ band-merging is simply $\EE = \log_2 2^n = n$.
That is, the process carries $n$ bits of phase information. Putting
these facts together, then, we have a very simple relationship in
the complexity-entropy diagram at band-mergings:
\begin{equation}
  \EE \, = \, -\log_2 \hmu ~.
\label{tent.theory}
\end{equation}
This is graphed as the dashed line in Fig.~\ref{tent.plot}.
It is clear that the entire complexity-entropy diagram is much richer
than this simple expression indicates. Nonetheless, Eq.
(\ref{tent.theory}) does capture the overall shape quite well.

Note that, in sharp contrast to the logistic map, for the tent map it
does appear as if the excess entropy takes on only a single value for each
value of the entropy rate $\hmu$. The reason for this is straightforward.
The entropy rate $\hmu$ is a simple monotonic function of the parameter
$a$---$h_\mu = \log_2 a$---and so there is a one-to-one relationship
between them. As a result, each $h_\mu$ value on the complexity-entropy
diagram corresponds to one and only one value of $a$ and, in turn, corresponds
to one and only one value of $\EE$. Interestingly, the excess entropy appears
to be a continuous function of $h_\mu$, although not a differentiable one. 

\subsection{Ising Spin Systems}

We now investigate the complexity-entropy diagrams of the Ising model
in one and two spatial dimensions. Ising models are among the simplest
physical models of spatially extended systems.  Originally introduced
to model magnetic materials, they are now used to model a wide range
of cooperative phenomena and order-disorder transitions and, more
generally, are viewed as generic models of spatially extended,
statistical mechanical systems \cite{Chri05b,Seth06a}.  Like the
logistic and tent maps, Ising models are also studied as an
intrinsically interesting mathematical topic. 
As we will see, Ising models provide an
interesting contrast with the intrinsic computation seen in the
interval maps.

Specifically, we consider spin-$1/2$ Ising models
with nearest (NN) and next-nearest neighbor (NNN) interactions. The
Hamiltonian (energy function) for such a system is:
\begin{eqnarray}
  {\cal H} \,& =&  \, -J_1 \sum_{\langle i,j\rangle_{\rm nn}} S_{i} S_{j}
  \nonumber \\ & &  -J_2
  \sum_{\langle i,j\rangle_{{\rm nnn}} } S_{i} S_{j} \,-\, B \sum_{i} S_{i} \;,
\label{Hamiltonian}
\end{eqnarray}
where the first (second) sum is understood to run over all NN (NNN)
pairs of spins.  In one dimension, a spin's nearest-neighbors will
consist of two spins, one to the right and one to the left, whereas
in two dimensions a spin will have four nearest neighbors---left,
right, up, and down.  Each spin $S_{i}$ is a binary variable: $S_{i}
\,\in\, \{-1,+1\}$.  The coupling constant $J_1$ is a parameter that
when positive (negative) makes it energetically favorable for NN
spins to (anti-)align.  The constant $J_2$ has the same effect on NNN
spins.  The parameter $B$ may be viewed as an external field; its
effect is to make it energetically favorable for spins to point up
(i.e., have a value of $+1$) instead of down.  The probability of a
configuration is taken to be proportional to its Boltzmann weight:
the probability of a spin configuration ${\cal C}$ is proportional
to $e^{-\beta{\cal H}({\cal C})}$, where $\beta = 1/T$ is the inverse
temperature.

In equilibrium statistical mechanics, the entropy density is a
monotonic increasing function of the temperature.  Quite generically,
a plot of the entropy $h_\mu$ as a function of temperature $T$
resembles that of the top plot in Fig.~\ref{2DCriticalhvsE}.  Thus,
$h_\mu$ may be viewed as a nonlinearly rescaled temperature.  One
might ask, then, why one might want to plot complexity versus entropy:
Isn't a plot of complexity versus temperature qualitatively the same?
Indeed, the two plots would look very similar.  However, there
are two major benefits of complexity-entropy diagrams for statistical
mechanical systems.  First,
the entropy captures directly the system's unpredictability, measured
in bits per spin.  The entropy thus measures the system's
information processing properties.  Second, plotting complexity versus
entropy and not temperature allows for a direct comparison of the
range of information processing properties of statistical mechanical
systems with systems for which there is not a well defined
temperature, such as the deterministic dynamical systems of the
previous section or the cellular automata of the subsequent one.

\subsubsection{One-Dimensional Ising System}

\label{1D.spin.section}

\begin{figure}[tbp]
\epsfxsize=3.5in
\begin{center}
\leavevmode
\epsffile{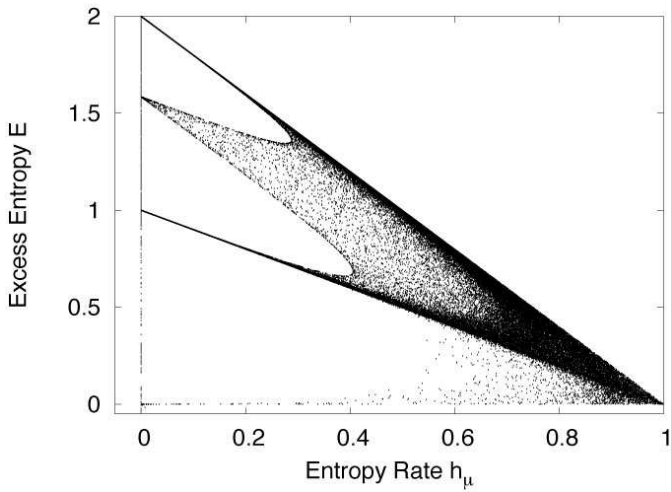}
\end{center}
\vspace{-6mm}
\caption{Complexity-entropy diagram for the one-dimensional, spin-$1/2$
  antiferromagnetic Ising model with nearest- and next-nearest-neighbor
  interactions. $10^5$ system parameters were sampled randomly from the
  following ranges: $J_1 \in [-8,0]$, $J_2 \in [-8,0]$,
  $T \in [0.05,6.05]$, and $B \in [0,3]$. For each parameter setting,
  the excess entropy $\EE$ and entropy density $\hmu$ were calculated
  analytically.
  }
\label{Ising.Batcape}
\vspace{-2mm}
\end{figure}

We begin by examining one-dimensional Ising systems.
In Refs.~\cite{Crut97a,Feld98b,Feld98c} two of the authors developed
exact, analytic transfer-matrix methods for calculating $\hmu$ and
$\EE$ in the thermodynamic ($N \rightarrow \infty$) limit.  These
methods make use of the fact the NNN Ising model is order-$2$
Markovian.  We used
these methods to produce Fig.~\ref{Ising.Batcape}, the
complexity-entropy diagram for the NNN Ising system with
antiferromagnetic coupling constants $J_1$ and $J_2$ that tend to
anti-align coupled spins. The figure gives a
scatter plot of $10^5$ $(\hmu, \EE)$ pairs for system parameters that
were sampled randomly
from the following ranges: $J_1 \in [-8,0]$, $J_2 \in [-8,0]$, $T \in
[0.05,6.05]$, and $B \in [0,3]$. For each parameter realization, the
excess entropy $\EE$ and entropy density $\hmu$ were calculated.
Fig.~\ref{Ising.Batcape} is rather striking---the $(\hmu,\EE)$ pairs
are organized in the shape of a ``batcape''. Why does the plot have
this form?

Recall that if a sequence over a binary alphabet is periodic with period
$p$, then $\EE = \log_2 p$ and $\hmu = 0$. Thus, the ``tips'' of the batcape
at $\hmu = 0$ correspond to crystalline (periodic) spin configurations with
periods $1$, $2$, $3$, and $4$. For example, the $(0,0)$ point is the
period-$1$ configuration with all spins aligned. These periodic regimes
correspond to the system's different possible ground states. As the entropy
density increases, the cape tips widen and eventually join.

Figure~\ref{Ising.Batcape} demonstrates in graphical form that there
is organization in the process space defined by the Hamiltonian of
Eq.~(\ref{Hamiltonian}). Specifically, for antiferromagnetic couplings,
$\EE$ and $\hmu$ values do not uniformly fill the plane. There are forbidden
regions in the complexity-entropy plane. Adding randomness ($\hmu$) to
the periodic ground states does not immediately destroy them.  That is,
there are low-entropy states that are almost-periodic. The apparent upper
linear bound is that of Eq.~(\ref{EE_hmu_bound}) for a system with
at most $4$ Markov states or, equivalently, a order-$2$ Markov chain:
$\EE \leq 2 - 2 \hmu$.

In contrast, in the logistic map's complexity-entropy diagram
(Fig.~\ref{logistic.banded.plot}) one does not see anything
remotely like the batcape. This indicates that there are no
low-entropy, almost-periodic configurations related to the exactly
periodic configurations generated at zero-entropy along the
period-doubling route to chaos. Increasing the parameter there
does not add randomness to a periodic orbit. Rather, it causes a
system bifurcation to a higher-period orbit.

\subsubsection{Two-Dimensional Ising Model}

Thus far we have considered only one-dimensional systems, either
temporal or spatial.  However,
the excess entropy can be extended to apply to two-dimensional
configurations as well; for details, see Ref.~\cite{Feld03a}.  Using
methods from there, we calculated the excess entropy
and entropy density for the two-dimensional Ising model with nearest-
and next-nearest-neighbor interactions.  In other words, we calculated
the complexity-entropy diagram for the two-dimensional version of the
system whose complexity-entropy diagram is shown in
Fig.~\ref{Ising.Batcape}.  There are several different definitions for
the excess entropy in two dimensions, all of which are similar but not
identical.  In Fig.~\ref{Ising.Batcape} we used a version that is
based on the mutual information and, hence, is denoted ${\EE}_I$
\cite{Feld03a}.

\begin{figure}[tbp]
\epsfxsize=3.5in
\begin{center}
\leavevmode
\epsffile{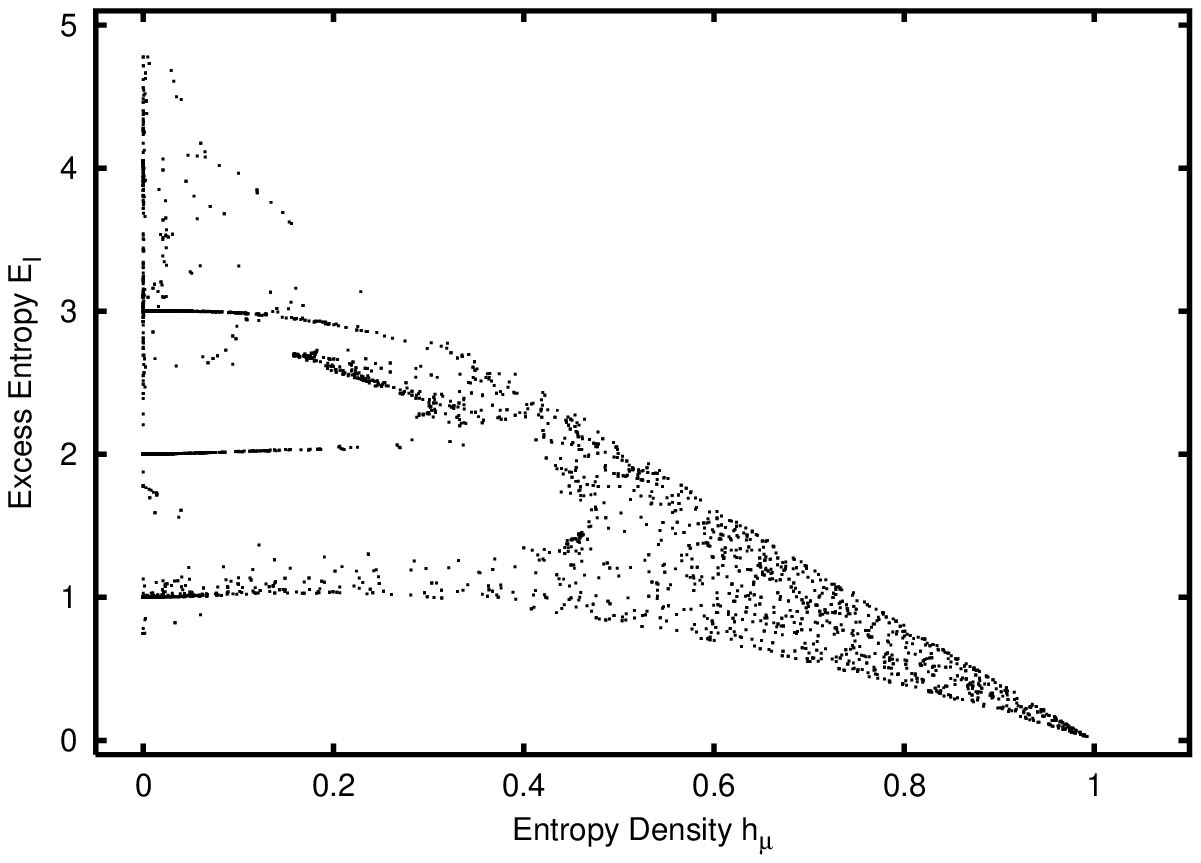}
\end{center}
\vspace{-6mm}
\caption{Complexity-entropy diagram for the two-dimensional, spin-$1/2$
  anti\-ferromagnetic Ising model with nearest- and next-nearest-neighbor
  interactions. System parameters were sampled randomly from the
  following ranges: $J_1 \in [-3,0]$, $J_2 \in [-3,0]$, $T \in
  [0.05,4.05]$, and $B = 0$. For each parameter setting, the excess
  entropy $\EE_I$ and entropy density $\hmu$ were estimated
  numerically.}
\vspace{-4mm}
\label{2DIsingBatcape}
\end{figure}

Figure \ref{2DIsingBatcape} gives a scatter plot of $4,500$
complexity-entropy pairs.  System parameters in
Eq.~(\ref{Hamiltonian}) were sampled randomly from the following
ranges: $J_1 \in [-3,0]$, $J_2 \in [-3,0]$, $T \in [0.05,4.05]$, and
$B = 0$. For each parameter setting, the excess entropy $\EE_I$ and
entropy density $\hmu$ were estimated numerically; the configurations
themselves were generated via a Monte Carlo simulation. For each $(\hmu,\EE)$
point the simulation was run for $200,000$ Monte Carlo updates per site to
equilibrate. Configuration data was then taken for $20,000$ Monte Carlo
updates per site. The lattice size was a square of $48 \times 48$ spins.
The long equilibration time is necessary because, for some Ising models at
low temperature, single-spin flip dynamics of the sort used here have very
long transient times \cite{Spir01a,Spir01b,Vazq02a}.

Note the similarity between Figs.~\ref{Ising.Batcape} and \ref{2DIsingBatcape}.
For the 2D model, there is also a near-linear upper bound:
$\EE \leq 5 (1-h_\mu)$. In addition, one sees periodic
spin configurations, as evidenced by the horizontal bands. An $\EE_I$ of
$1$ bit corresponds to a checkerboard of period $2$; $\EE_I = 3$
corresponds to a checkerboard of period $4$; while $\EE_I = 2$
corresponds to a ``staircase'' pattern of period $4$.  See
Ref.~\cite{Feld03a} for illustrations.  The two period-$4$
configurations are both ground states for the model in the parameter
regime in which $|J_2| < |J_1|$ and $J_2 < 0$. At low temperatures,
the state into which the system settles is a matter of chance.

Thus, the horizontal streaks in the low-entropy region of
Fig.~\ref{2DIsingBatcape} are the different ground states possible for
the system.  In this regard Fig.~\ref{2DIsingBatcape} is qualitatively
similar to Fig.~\ref{Ising.Batcape}---in each there are several
possible ground states at $\hmu = 0$ that persist as the entropy
density is increased. However, in the two-dimensional system of
Fig.~\ref{2DIsingBatcape} one sees a scatter of other values
around the periodic bands.  There are even $\EE_I$ values larger than
$3$.  These $\EE_I$ values arise when parameters are selected in which
the NN and NNN coupling strengths are similar; $J_1 \approx J_2$.
When this is the case, there is no energy cost associated with a
horizontal or vertical defect between the two possible ground states.
As a result, for low temperatures the systems effectively freezes into
horizontal or vertical strips consisting of the different ground states.
Depending on the number of strips and their relative widths, a number
of different $\EE_I$ values are possible, including values well above
$3$, indicating very complex spatial structure.

Despite these differences, the similarities between the
complexity-entropy plots for the one- and two-dimensional systems is
clearly evident.  This is all the more noteworthy since one- and
two-dimensional Ising models are regarded as very different sorts of
system by those who focus solely on phase transitions. The
two-dimensional Ising model has a critical phase transition while the
one-dimensional does not.  And, more generally, two-dimensional random
fields are generally considered very different mathematical entities
than one-dimensional sequences. Nevertheless, the two complexity-entropy
diagrams show that, away from criticality, the one- and two-dimensional
Ising systems' ranges of intrinsic computation are similar.

\subsubsection{Ising Model Phase Transition}

As noted above, the two-dimensional Ising
model is well known as a canonical model of a system that undergoes a
continuous phase transition---a discontinuous change in the system's
properties as a parameter is continuously varied.  The 2D NN Ising
model with ferromagnetic ($J_1>0$) bonds and no NNN coupling ($J_2 =
0$) and zero external field ($B=0$)
undergoes a phase transition at $T = T_c \approx 2.269$ when $J_1 =
1$.  At the critical temperature $T_c$ the magnetic susceptibility
diverges and the specific heat is not differentiable.  In
Fig.~\ref{2DIsingBatcape} we restricted ourselves to antiferromagnetic
couplings and thus did not sample in the region of parameter space in
which the phase transition occurs.

\begin{figure}[tbp]
\epsfxsize=3.5in
\begin{center}
\leavevmode
\epsffile{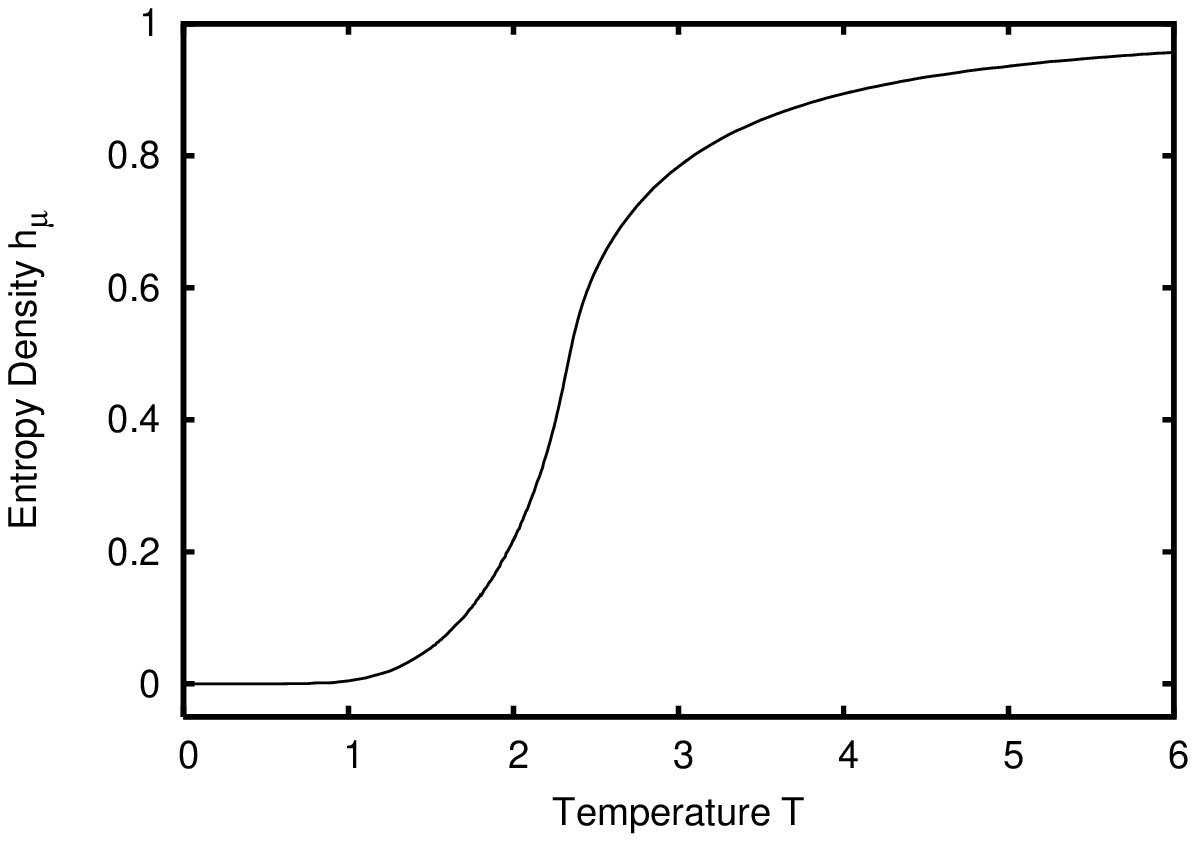}\\
\vspace{-5mm}
\epsfxsize=3.5in
\epsffile{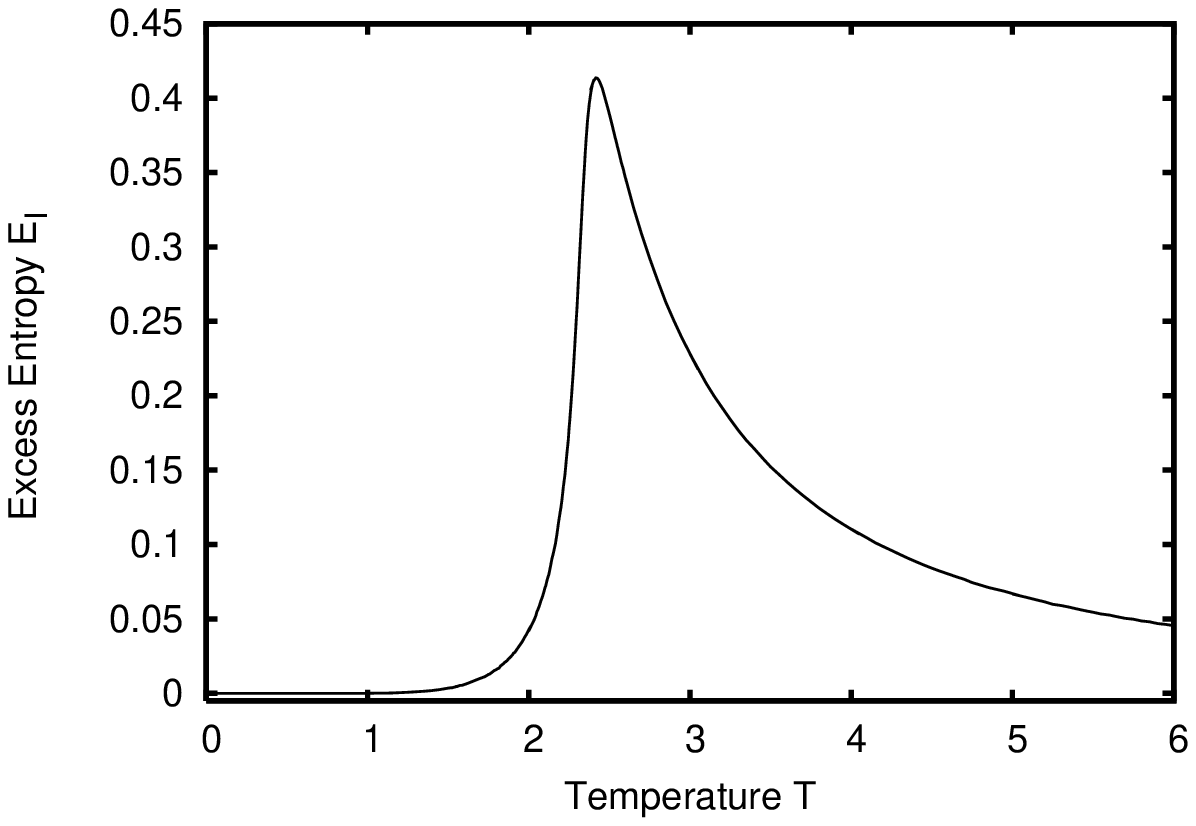}\\
\vspace{-5mm}
\epsfxsize=3.5in
\epsffile{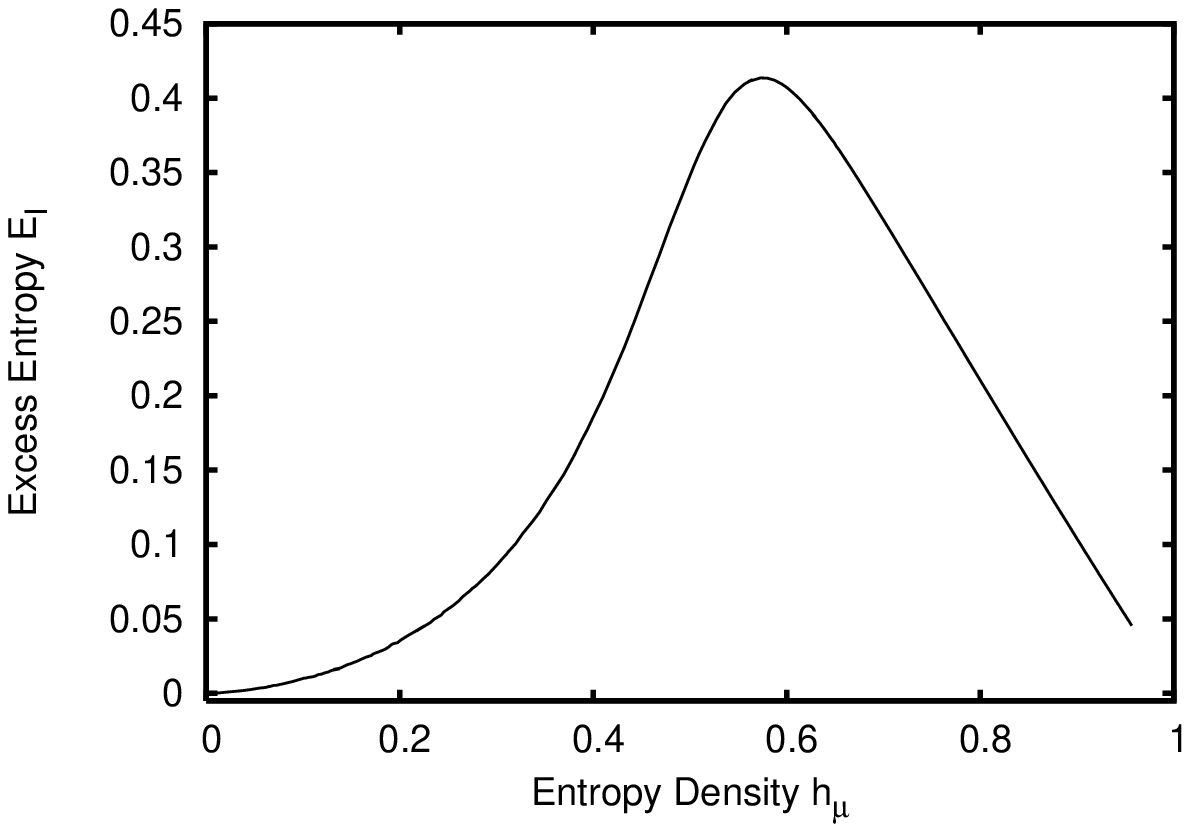}
\end{center}
\vspace{-6mm}
\caption{Entropy rate vs.~temperature, excess entropy vs.~temperature, and
  the complexity-entropy diagram for the 2D NN ferromagnetic Ising
  model.  Monte Carlo results for $200$ temperatures between $0$ and
  $6$.  The temperature was sampled more densely near the critical
  temperature. For further discussion, see text.  }
\vspace{-4mm}
\label{2DCriticalhvsE}
\end{figure}

What happens if we fix $J_1 = 1$, $J_2 = 0$, and $B=0$, and vary the
temperature? In this case, we see that the complexity, as measured by $\EE$,
shows a sharp maximum near the critical temperature $T_c$. Figure
\ref{2DCriticalhvsE} shows results obtained via a Monte Carlo simulation on
a $100 \times 100$ lattice. We used a Wolff cluster algorithm and periodic
boundary conditions. After $10^6$ Monte Carlo steps (one step is one proposed
cluster flip), $25,000$ configurations were sampled, with $200$ Monte
Carlo steps between measurements.  This process was repeated for over
$200$ samples between $T=0$ and $T=6$.  More temperatures were sampled
near the critical region.  

In Fig.~\ref{2DCriticalhvsE} we first plot entropy density $\hmu$ and
excess entropy $\EE$ versus temperature.  As expected, the
excess entropy reaches a maximum at the critical temperature $T_c$.
At $T_c$ the correlations in the system decay algebraically, whereas
they decay exponentially for all other $T_c$ values.  Hence, $\EE$,
which may be viewed as a global measure of correlation, is maximized at
$T_c$.  For the system of Fig.~\ref{2DCriticalhvsE}, $T_c$ appears to
have an approximate value of $2.42$. This is above the exact value for an
infinite system, which is $T_c \approx 2.27$. Our estimated value is higher,
as one expects for a finite lattice. At the critical temperature,  $h_\mu
\approx 0.57$, and $\EE \approx 0.413$.

Also in Fig.~\ref{2DCriticalhvsE} we show the complexity-entropy
diagram for the 2D Ising model.  This complexity-entropy diagram is a
single curve, instead of the scatter plots seen in the
previous complexity-entropy diagrams.  The reason is that
we varied a single parameter, the temperature, and
entropy is a single-valued function of the temperature, as can clearly
be seen in the first plot in Fig.~\ref{2DCriticalhvsE}.  Hence, there
is only one value of $\hmu$ for each temperature, leading to a single
curve for the complexity-entropy diagram.

Note that the peak in the complexity-entropy diagram for the
2D Ising model is rather rounded, whereas $\EE$ plotted versus
temperature shows a much sharper peak.  The reason for this rounding
is that the entropy density $\hmu$ changes very rapidly near $T_c$.
The effect is to smooth the $\EE$ curve when plotted against $\hmu$.

A similar complexity-entropy was produced by Arnold \cite{Arno96}. He also
estimated the excess entropy, but did so by considering only one-dimensional
sequences of measurements obtained at a single site, while a Monte Carlo
simulation generated a sequence of two-dimensional configurations. Thus,
those results do not account for two-dimensional structure but, rather,
reflect properties of the dynamics of the particular Monte Carlo updating
algorithm used. Nevertheless, the results of Ref.~\cite{Arno96} are
qualitatively similar to ours. 

Erb and Ay \cite{Erb04a} have calculated the \emph{multi-information} for
the two-dimensional Ising model as a function of temperature. The
multi-information is the difference between the entropy rate and the
entropy of a single site: $H(1) - h_\mu$. That is, the multi-information
is only the leading term in the sum which defines the excess entropy,
Eq.~(\ref{E.def}). (Recall that $h_\mu(1) = H(1)$.) They find that the
multi-information is a continuous function of the temperature and that
it reaches a sharp peak at the critical temperature \cite[Fig.~4]{Erb04a}. 

\subsection{Cellular Automata}

The next process class we consider is \emph{cellular automata} (CAs)
in one and two spatial dimensions. Like spin systems, CAs are common
prototypes used to model spatially extended dynamical systems. For reviews
see, e.g., Refs.~\cite{Wolf83a,Chop98a,Ilac01a}. Unlike the Ising
models of the previous section, the CAs that we study here are
deterministic. There is no noise or temperature in the system. 

The states of the CAs we shall consider consist of one- or
two-dimensional \emph{configurations}
${\mathbf s} = \ldots s^{-1} , s^0 , s^1 , \ldots $ of discrete $K$-ary
\emph{local states} $s^i \in \{ 0, 1, \ldots , K-1 \}$. The
configurations change in time according to a \emph{global update
function} $\mathbf \Phi$:
\begin{equation}
{\mathbf s}_{t+1}^i = {\mathbf \Phi} {\mathbf s}_t^i ~,
\end{equation}
starting from an \emph{initial configuration} ${\mathbf s}_0$. What makes CAs
\emph{cellular} is that configurations evolve according to a
\emph{local update rule}. The value $s_{t+1}^i$ of site $i$ at the next time
step is a function $\phi$ of the site's previous value and the values of
neighboring sites within some \emph{radius} $r$:
\begin{equation}
s_{t+1}^i = \phi ( s_t^{i-r} \ldots, s_t^i \ldots, s_t^{i+r} ) ~.
\end{equation}
All sites are updated synchronously. The CA update rule $\phi$ consists
of specifying the \emph{output value} $s_{t+1}$ for all possible
\emph{neighborhood configurations}
$\eta_t = s_t^{i-r} \ldots, s_t^i \ldots, s_t^{i+r}$.
Thus, for 1D radius-$r$ CAs, there are $K^{2r+1}$ possible neighborhood
configurations and $2^{K^{2r+1}}$ possible CA rules. The $r=1$, $K=2$
1D CAs are called {\em elementary cellular automata}\/ \cite{Wolf83a}.

In all CA simulations reported we began with an arbitrary random initial
configuration ${\mathbf s}_0$ and iterated the CA several thousand times
to let transient behavior die away. Configuration statistics were then
accumulated for an additional period of thousands of time steps, as
appropriate. Periodic boundary conditions on the underlying lattice
were used.



\begin{figure}[tbp]
\epsfxsize=3.0in
\begin{center}
\leavevmode
\epsffile{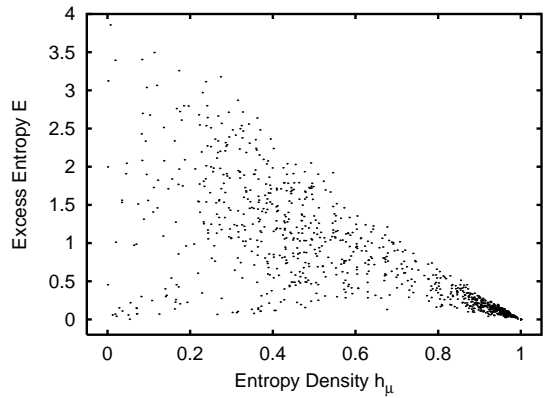}
\end{center}
\vspace{-6mm}
\caption{Spatial entropy density $h^s_\mu$ and spatial excess entropy
  $\EE^s$ for a random sampling of $10^3$ $r = 2$, binary 1D CAs.
   }
\vspace{-2mm}
\label{1D.rad2.spatial.hvsE}
\end{figure}

In Fig.~\ref{1D.rad2.spatial.hvsE} we
show the results of calculating various complexity-entropy diagrams
for 1D, $r = 2$, $K=2$ (binary) cellular automata.  There are $2^{2^5}
\approx 4.3 \times 10^9$ such CAs.  We cannot examine all $4.3$
billion CAs; instead we sample the space these CAs uniformly.
For the data of
Fig.~\ref{1D.rad2.spatial.hvsE}, the lattice has $5\times 10^4$ sites
and a transient time of $5 \times 10^4$ iterations was used. We plot
$\hmu$ versus $\EE$ for spatial sequences.  Plots for the temporal
sequences are qualitatively similar. There are several things to
observe in these diagrams.

One feature to notice in Fig.~\ref{1D.rad2.spatial.hvsE} is that no sharp
peak in the excess entropy appears at some intermediate $\hmu$ value. In
contrast, the maximum possible excess entropy falls off moderately rapidly
with increasing $\hmu$. A linear upper bound, $\EE \leq 4 ( 1 - h_\mu)$,
is almost completely respected. Note that, as is the case with the other
complexity-entropy diagrams presented here, for all $\hmu$ values except
$\hmu = 1$, there is a range of possible excess entropies.




In the early 1990's there was considerable exploration of the
organization of CA rule space. In particular, a series of papers
\cite{Lang90a,Li90b,Woot90a,Lang91a} looked at two-dimensional
eight-state ($K=8$) cellular automata, with a neighborhood size of
$5$ sites---the site itself and its nearest neighbor to the north,
east, west, and south. These references reported evidence for
the existence of a phase transition in the complexity-entropy diagram at a
critical entropy level. In contrast, however, here and in the previous
sections we find no evidence for such a transition. The reasons that
Refs.~\cite{Lang90a,Li90b,Woot90a,Lang91a} report a transition are two-fold.
First, they used very restricted measures of randomness and complexity:
entropy of \emph{single isolated} sites and mutual information of neighboring
\emph{pairs} of single sites, respectively. These choices have the effect of
projecting organization \emph{onto} their complexity-entropy diagrams. The
organization seen is largely a reflection of constraints on the chosen
measures, not of intrinsic properties of the CAs. Second, they do not sample
the space of CA's uniformly; rather, they parametrize the space of CAs and
sample only by sweeping their single parameter. This results in a sample of
CA space that is very different from uniform and that is biased toward higher
complexity CAs. For a further discussion of complexity-entropy diagrams for
cellular automata, including a discussion of
Refs.~\cite{Lang90a,Li90b,Woot90a,Lang91a}, see Ref.~\cite{Feld06a}.

\subsection{Markov Chain Processes}

In this and the next section, we consider two classes of process that
provide a basis of comparison for the preceding nonlinear dynamics
and statistical mechanical systems: those generated by Markov chains and
topological \eMs. These classes are complementary to each other in the
following sense.  Topological \eMs\ represent structure in terms of which
sequences (or configurations) are allowed or not. When we explore the space
of topological \eMs, the associated processes differ in which sets of
sequences occur and which are forbidden. In contrast, when exploring
Markov chains, we fix a set of allowed words---in the present case the
full set of binary sequences---and then vary the probability with
which subwords occur. These two classes thus represent two different
types of possible organization in intrinsic computation---types that
were mixed in the preceding example systems.  

\begin{figure}[tbp]
\epsfxsize=3.0in
\begin{center}
\leavevmode
\epsffile{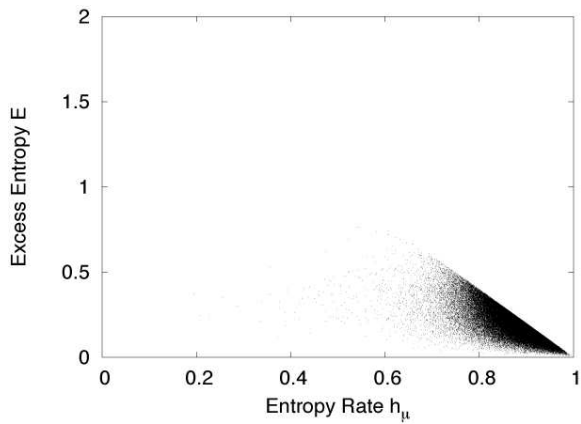}
\end{center}
\vspace{-6mm}
\caption{Excess-entropy, entropy-rate pairs for $10^5$ randomly
  selected $4$-state Markov chains.
  }
\vspace{-2mm}
\label{null}
\end{figure}

In Fig.~\ref{null} we plot $\EE$ versus $\hmu$ for order-$2$
($4$-state) Markov chains over a binary alphabet. Each element in the
stochastic transition matrix $T$ is chosen uniformly from the unit
interval. The elements of the matrix are then normalized row by row so
that $\sum_j T_{ij} = 1$. We generated $10^5$ such matrices and formed
the complexity-entropy diagram shown in Fig.~\ref{null}. Since these
processes are order-$2$ Markov chains, the bound of
Eq. (\ref{EE_hmu_bound}) applies.  This bound is the sharp, linear
upper limit evident in Fig.~\ref{null}: $\EE \, = 2 - 2\hmu$. 

It is illustrative to compare the $4$-state Markov chains considered
here with the 1D NNN Ising models of Sec. \ref{1D.spin.section}.  
The order-$2$ (or $4$-state Markov) chains with a binary alphabet are
those systems
for which the value of a site depends on the previous two sites, but
no others.  In terms of spin systems, then, this is a spin-$1/2$
(i.e., binary) system with nearest- and  next-nearest neighbors.  The
transition matrix for the Markov chain is $4 \times 4$ and thus has
$16$ elements.  However, since each row of the transition matrix
must be normalized, there are $12$ independent parameters for this
model class.  In contrast, there are only $3$ independent parameters
for the 1D NNN Ising chain---the parameters $J_1$, $J_2$, $B$, and
the temperature $T$.  One of the parameters may be viewed as
setting an energy scale, so only three are independent.

Thus, the 1D NNN systems are a proper subset of the $4$-state Markov chains.
Note that their complexity-entropy diagrams are very different, as a quick 
glance at Figs.~\ref{Ising.Batcape} and \ref{null} confirms. The reason for
this is that the Ising model, due to its parametrization (via the Hamiltonian
of Eq.~(\ref{Hamiltonian})), samples the space of processes in a
very different way than the Markov chains. This underscores the
crucial role played by the choice of model and, so too, the choice in
parametrizing a model space. Different parametrizations of the same
model class, when sampled uniformly over those parameters, yield
complexity-entropy diagrams with different structural properties.

\subsection{The Space of Processes: Topological \EMs}

The preceding model classes are familiar from dynamical systems theory,
statistical mechanics, and stochastic process theory. Each has served
an historical purpose in their respective fields---purposes that reflect
mathematically, physically, or statistically useful parametrizations of
the space of processes. In the preceding sections we explored these
classes, asked what sort of processes they could generate, and then
calculated complexity-entropy pairs for each process to reveal the
range of possible information processing within each class.

Is there a way, though, to directly explore the space of processes, without
assuming a particular model class or parametrization? Can each process be
taken at face value and tell us how it is structured? More to the point, can
we avoid making structural assumptions, as done in the preceding sections?

Affirmative answers to these questions are found in the approach laid out
by \emph{computational mechanics} \cite{Crut89,Crut92c,Shal01a}.
Computational mechanics demonstrates that each process has an optimal,
minimal, and unique representation---the \emph{\eM}---that captures the
process's structure. Due to optimality, minimality, and uniqueness, the
\eM\ may be viewed as \emph{the} representation of its associated process. In
this sense, this representation is parameter free. To determine an \eM\ for a
process one calculates a set of \emph{causal states} and their transitions.
In other words, one does not specify a priori the number of states or
the transition structure between them. Determining the \eM\ makes such no
structural assumptions \cite{Crut92c,Shal01a}.  

Using the one-to-one relationship between processes and their \eMs, here we
invert the preceding logic of going from a process to its \eM. We explore the
space of processes by systematically enumerating \eMs\ and then calculating
their excess entropies ${\bf E}$ and their entropy rates $h_\mu$. This gives
a direct view of how intrinsic computation is organized in the space of
processes.

As a complement to the Markov chain exploration of how intrinsic computation
depends on transition probability variation, here we examine how an \eM's
structure (states and their connectivity) affects information processing. We
do this by restricting attention to the class of \emph{topological \eMs} whose
branching transition probabilities are fair (equally probable). (An example
is shown in Fig.~\ref{f_mn}.)

If we regard two \eMs\ isomorphic up to variation in transition
probabilities as members of a single equivalence class, then each such
class of \eMs\ contains precisely one topological \eM. (Symbolic dynamics
\cite{Lind95a} refers to a related class of representations as
\emph{topological Markov chains}. An essential, and important,
difference is that \eMs\ always have the smallest number of states.)

\begin{table}
\begin{tabular}{|c|c|}
\hline
  {Causal States} & {Topological}   \\
       n          & {\eMs}          \\
\hline
       1          & 3               \\
       2          & 7               \\
       3          & 78              \\
       4          & 1,388           \\
       5          & 35,186          \\
\hline
\end{tabular}
\caption{The number of topological binary \eMs\ up to $n = 5$ causal states.
  (After Ref. \protect\cite{McTa05a}.)
\label{proclangcount}
  }
\end{table}

It turns out that the topological \eMs\ with a finite number of states can be
systematically enumerated \cite{McTa05a}. Here we consider only \eMs\ for
binary processes: $\mathcal{A} = \{0,1\}$. Two \eMs\ are
isomorphic and generate essentially the same stochastic process,
if they are related by a relabeling of states or if their output
symbols are exchanged: $0$ is mapped to $1$ and vice versa. The number
of isomorphically distinct topological \eMs\ of $n=1,\ldots,5$~states
is listed in Table~\ref{proclangcount}.

\begin{figure*}[tbp]
\epsfxsize=6.0in
\begin{center}
\leavevmode
\epsffile{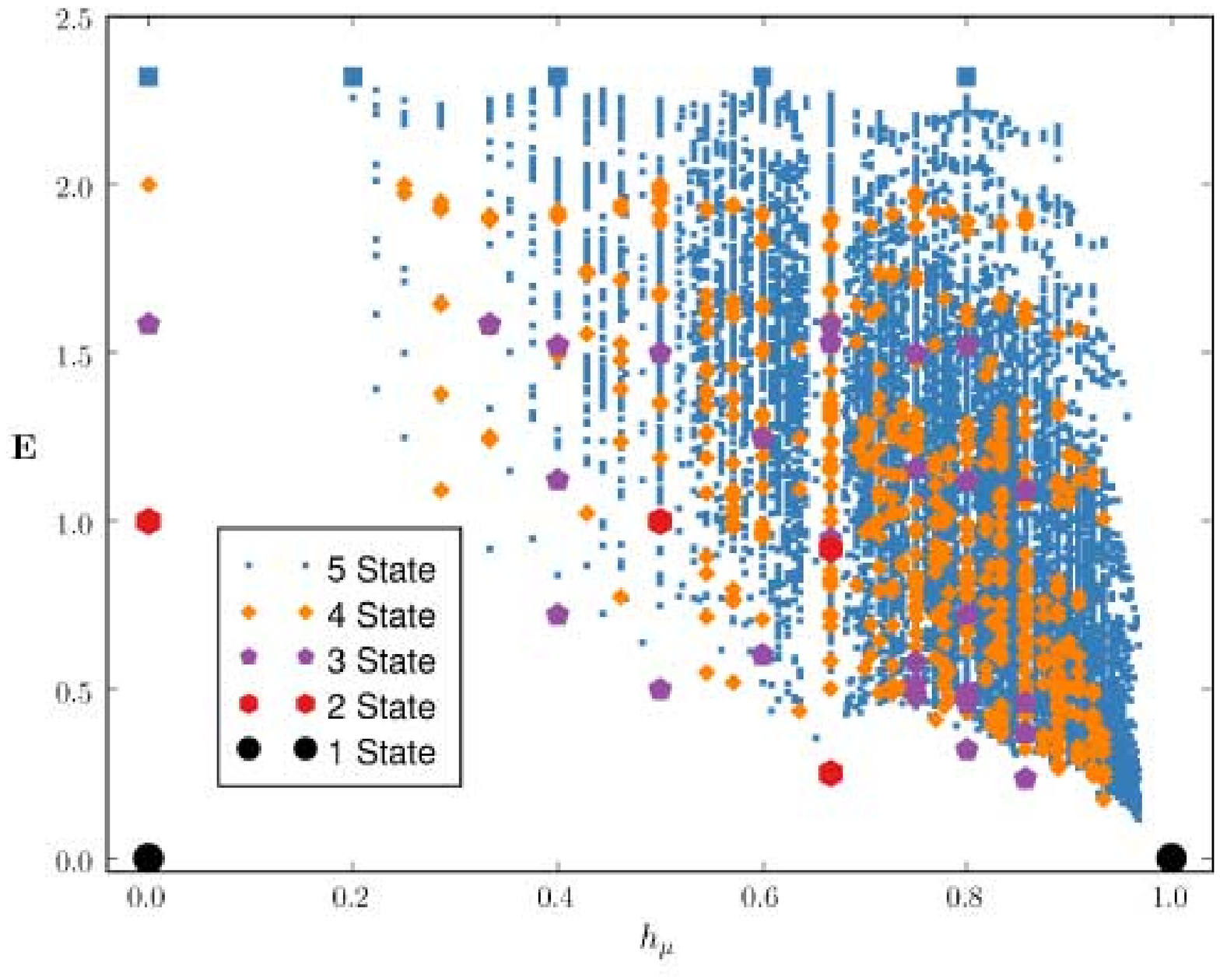}
\end{center}
\vspace{-6mm}
\caption{Complexity-entropy pairs $(h_\mu,\EE)$ for all topological binary
  \eMs\ with $n = 1, \ldots, 4$ states and for $35,041$ of the $35,186$
  $5$-state \eMs. The excess entropy is estimated as $\EE(L) = H(L) - L \hmu$
  using the exact value for the entropy rate $\hmu$ and a storage-efficient
  type-class algorithm \cite{Youn93a} for the block entropy $H(L)$. The
  estimates were made by increasing $L$ until $\EE(L) - \EE(L-1) < \delta$,
  where $\delta = 0.0001$ for $1$, $2$, and $3$ states; $\delta = 0.0050$
  for $4$ states; and $\delta = 0.0100$ for 5 states.
  }
\label{process.plot}
\end{figure*}

In Fig.~\ref{process.plot} we plot their $(\hmu,\EE)$ pairs. There one
sees that the complexity-entropy diagram exhibits quite a bit of
organization, with variations from very low to very high density of
\eMs\ co-existing with several distinct vertical (iso-entropy)
families. To better understand the structure in the complexity-entropy
diagram, though, it is helpful to consider bounds on the complexities
and entropies of Fig.~\ref{process.plot}. The minimum complexity, $\EE = 0$,
corresponds to machines with only a single state. There are two possibilities
for such binary \eMs. Either they generate all $1$s (or $0$s) or all sequences
occurring with equal probability (at each length). If the latter, then
$\hmu = 1$; if the former, $\hmu = 0$. These two points, $(0,0)$ and $(1,0)$,
are denoted with solid circles along Fig.~\ref{process.plot}'s horizontal axis.

\begin{figure}[tbp]
\epsfxsize=3.0in
\begin{center}
\leavevmode
\epsffile{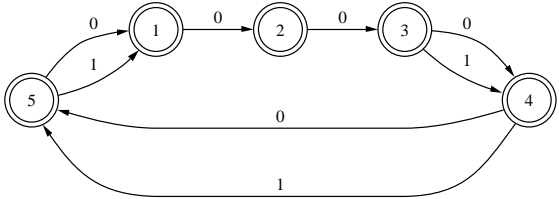}
\end{center}
\caption{An example topological \eM\ for a cyclic process in $\mc{F}_{5,3}$.
  Note that branching occurs only between pairs of successive states in
  the cyclic chain.   The excess entropy for this process is $\log_2 5
  \approx 2.32$, and the entropy rate is $3/5$. 
  }
\label{f_mn}
\end{figure}

The maximum $\EE$ in the complexity-entropy diagram is $\log_2 5
\approx 2.3219$.  One such \eM\ corresponds to the zero-entropy,
period-$5$ processes. And there are four similar processes with
periods $p = 1, 2, 3, 4$ at the points $(0, \log_2 p)$. These are
denoted on the figure by the tokens along the left vertical
axis. 

There are other period-$5$ \emph{cyclic, partially random} processes with
maximal complexity, though; those with causal states in a cyclic
chain. These have $b = 1, 2, 3, 4$ branching transitions between successive
states in the chain and so positive entropy. These appear as a horizontal
line of enlarged square tokens along in the upper portion of the
complexity-entropy diagram.
Denote the family of $p$-cyclic processes with $b$ branchings as
$\mc{F}_{p,b}$. An \eM\ illustrating $\mc{F}_{5,3}$ is shown in
Fig.~\ref{f_mn}.  The excess entropy for this process is $\log_2 5
\approx 2.32$, and the entropy rate is $3/5$. 

Since \eMs\ for cyclic processes consist of states in a single loop,
their excess entropies provide an upper bound among \eMs\ that generate
$p$-cyclic processes with $b$ branchings states, namely:
\begin{equation}
  \EE(\mc{F}_{p,b}) = \log_2 (p) ~.
\end{equation}
Clearly, $\EE(\mc{F}_{p,b}) \rightarrow \infty$ as $p \rightarrow \infty$.
Their entropy rates are given by a similarly simple expression:
\begin{equation}
  \hmu(\mc{F}_{p,b}) = \frac{b}{p} ~.
\end{equation}
Note that $\hmu(\mc{F}_{p,b}) \rightarrow 0$ as $p \rightarrow \infty$
with fixed $b$ and $\hmu(\mc{F}_{p,b}) \rightarrow 1$ as $b \rightarrow p$.
Together, then, the family $\mc{F}_{5,b}$ gives an upper bound to the
complexity-entropy diagram.

The processes $\mc{F}_{p,b}$ are representatives of the highest points
of the prominent jutting vertical towers of \eMs\ so prevalent in
Fig.~\ref{process.plot}. It therefore seems reasonable
to expect the $(\hmu,\EE)$~coordinates for $p$-cyclic process
languages to possess at least $p-1$~vertical towers, distributed
evenly at $\hmu=b/p$, $b=1, \dots, p-1$, and for these towers to
correspond with towers of $m$-cyclic process languages whenever $m$ is
a multiple of~$p$.

These upper bounds are one key difference from earlier classes in which there
was a decreasing linear upper bound on complexity as a function of entropy
rate: $\EE \leq R(1-h_\mu)$. That is, in the space of processes, many are not
so constrained. The subspace of topological \eMs\ illustrates that there are
many highly entropic, highly structured processes. Some of the more
familiar model classes appear to inherit, in their implied parametrization of
process space, a bias away from such processes.

It is easy to see that the families $\mc{F}_{p,p-1}$ and
$\mc{F}_{p,1}$ provide upper and lower bounds for~$\hmu$,
respectively, among the process languages that achieve
maximal~$\EE$ and for which~$\hmu > 0$. Indeed, the smallest
positive~$\hmu$ possible is achieved when only a single of
the equally probable states has more than one outgoing transition.

More can be said about this picture of the space of intrinsic
computation spanned by topological \eMs\ \cite{McTa05a}. Here,
however, our aim is to illustrate how rich the diversity of intrinsic
computation can be and to do so independent of conventional
model-class parametrizations. These results allow us to probe in a
systematic way a subset of processes in which structure dominates.

\section{Discussion and Conclusion}
\label{Discussion}

Complexity-entropy diagrams provide a common view of the intrinsic
information processing embedded in different processes. We used them
to compare markedly different systems: one-dimensional maps of the
unit interval; one- and two-dimensional Ising models; cellular automata;
Markov chains; and topological \eMs. The exploration of each class turned
different knobs in the sense that we adjusted different parameters:
temperature, nonlinearity, coupling strength, cellular automaton rule,
and transition probabilities.  Moreover, these parameters had very
different effects.  Changing the temperature and coupling constants in the
Ising models altered the probabilities of configurations, but it did not
change which configurations were allowed to occur. In contrast, the
topological \eMs\ exactly expressed what it means for different processes to
have different sets of allowed sequences. Changing the CA rules or the
nonlinearity parameter in the logistic map combined these effects: the
allowed sequences or the probability of sequences or both changed.
In this way, the survey illustrated in dramatic fashion one of the benefits
of the complexity-entropy diagram: it allows for a common comparison
across rather varied systems.  

For example, the complexity-entropy diagram for the radius-$2$,
one-dimensional cellular automata, shown in Fig.~\ref{1D.rad2.spatial.hvsE},
is very different from that of the logistic map, shown in
Fig.~\ref{logistic.banded.plot}.  For the logistic map, there is a
distinct lower bound for the excess entropy as a function of the
entropy rate. In Fig.~\ref{logistic.banded.plot} this is seen as the
large forbidden region at the diagram's lower portion. In sharp contrast,
in Fig.~\ref{1D.rad2.spatial.hvsE} no such forbidden region is seen.

At a more general level of comparison, the survey showed that for a given
$\hmu$, the excess entropy $\EE$ can be arbitrarily small.  This suggests
that the intrinsic computation of cellular automata and the logistic map are
organized in fundamentally different ways.  In turn, the 1D and 2D
Ising systems exhibit yet another kind of
information processing capability. Each of has well defined ground
states---seen as the zero-entropy tips of the ``batcapes'' in
Figs.~\ref{Ising.Batcape} and \ref{2DIsingBatcape}. These ground states are
robust under small amounts of noise---i.e., as the temperature increases from
zero. Thus, there are almost-periodic configurations at low entropy. In
contrast, there do not appear to be any almost-periodic configurations at low
entropy for the logistic map of Fig.~\ref{logistic.banded.plot}.

Our last example, topological \eMs, was a rather different kind of
model class. In fact, we argued that it gave a direct view into the
very structure of the space of processes. In this sense, the
complexity-entropy diagram was parameter free.  Note, however, that by
choosing all branching probabilities to be fair, we intentionally biased
this model class toward high-complexity, high-entropy processes. Nevertheless,
the distinction between the topological \eM\ complexity-entropy diagram of
Fig.~\ref{process.plot} and the others is striking.  

The diversity of possible complexity-entropy diagrams points to their
utility as a way to compare information processing across different classes.
Complexity-entropy diagrams can be empirically calculated from observed
configurations themselves. The organization reflected in the complexity-entropy
diagram then provides clues as to an appropriate model class to use for the
system at hand.  For example, if one found a complexity-entropy diagram with
a batcape structure like that of Figs.~\ref{Ising.Batcape} and
\ref{2DIsingBatcape}, this suggests that the class could be well modeled
using energies that, in turn, were expressed via a Hamiltonian.
Complexity-entropy diagrams may also be of use in classifying behavior
within a model class.  For example, as noted above, a type of
complexity-entropy diagram has already been successfully used to
distinguish between different types of structure in anatomical MRI
images of brains \cite{Youn05a,Youn08a}.  

Ultimately, the main conclusion to draw from this survey is that there
is a large diversity of complexity-entropy diagrams. There is certainly
not a universal complexity-entropy curve, as once hoped. Nor is it the case
that there are even qualitative similarities among complexity-entropy
diagrams.  They
capture distinctive structure in the intrinsic information processing
capabilities of a class of processes. This diversity is not a negative
result. Rather, it indicates the utility of this type of
intrinsic computation analysis, and it optimistically points
to the richness of information processing available in the mathematical and
natural worlds. Simply put, information processing is too complex to be
simply universal. 

\section*{Acknowledgments}

Our understanding of the relationships between complexity and entropy
has benefited from numerous discussions with Chris Ellison, Kristian
Lindgren, John Mahoney, Susan McKay, Cris Moore, Mats Nordahl, Dan Upper,
Patrick Yannul, and Karl Young. The authors thank, in particular, Chris
Ellison, for help in producing the \eM\ complexity-entropy diagram.
This work was supported at the Santa Fe Institute under the Computation,
Dynamics, and Inference Program via SFI's core grants from the National
Science and MacArthur Foundations. Direct support was provided from DARPA
contract F30602-00-2-0583. The CSC Network Dynamics Program funded by Intel
Corporation also supported this work. DPF thanks the Department of Physics
and Astronomy at the University of Maine for its hospitality. The REUs,
including one of the authors (CM), who worked on related parts of the
project at SFI were supported by the NSF during the summers of 2002 and
2003.

\bibliography{dpf}

\end{document}